\documentclass[a4paper]{jpconf}

\usepackage{graphicx}
\usepackage{citesort}
\usepackage[font=scriptsize]{caption}

\begin{document}

\title{Investigating gamma-ray fluxes from globular clusters}

\author{{H Ndiyavala$^{1,2}$, P Kr\"{u}ger$^{1}$, C Venter}$^{1}$\\
\begin{footnotesize}\normalfont
$^{1}${Centre for Space Research, North-West University, 11 Hoffman Street, Potchefstroom, 2531,
South Africa} \\
$^{2}${Department of Physics, University of Namibia, 380 Mandume Ndemufayo Avenue, Windhoek, 9000, Namibia}\\
E-mail: \texttt{hambeleleni.ndiyavala@gmail.com}
\end{footnotesize}}

\begin{abstract}
\noindent Globular clusters are large collections of old stars that are orbiting the core of a galaxy. Our Milky Way Galaxy has about 160 known clusters, with perhaps more to be discovered. We first accumulated the necessary parameters for 16 clusters and ran a numerical model that predicts the inverse Compton gamma-ray flux expected from each cluster. We also reanalysed data from 16 clusters observed by the H.E.S.S.\ very-high-energy (\textgreater100 GeV) gamma-ray telescopes. We confirmed the detection of Terzan~5 and found flux upper limits for the remaining 15 sources that were consistent with published results. We attempted to constrain some source parameters using   X-ray and gamma-ray data. We lastly list the five most promising clusters for future observations by the Cherenkov Telescope Array.
\end{abstract}

\section{Introduction}
\noindent Globular clusters (GCs) are among the most ancient bound stellar systems in the Universe. GCs are tight groups of $10^{4} - 10^{6}$~stars (e.g., \cite{Lang1992}). They are thought to have formed during the early stages of galaxy formation. GCs are spherically distributed about the Galactic Centre with an average distance of $\sim 12$~kpc. They contain exotic stellar members such as black holes, millisecond pulsars, white dwarfs, and cataclysmic variables.
The peculiar properties of these objects have been useful in diverse astrophysical disciplines such as cosmology, galaxy formation, stellar evolution, dynamics, as well as binary and variable stars \cite{Harris1996,Harris2010}.

The \textit{Fermi} Large Area Telescope (LAT), which is a gamma-ray satellite orbiting Earth, is continuously surveying the whole sky and has detected about a dozen GCs in the GeV band~\cite{Nolan2012}. On the other hand, the ground-based Cherenkov telescope H.E.S.S., which is operating in a pointing mode, has only plausibly detected a single GC within our Galaxy, namely Terzan~5~\cite{Abramowski2011}. Other Cherenkov telescopes could only produce upper limits in the very-high-energy (VHE) band for other Galactic GCs \cite{Anderhub2009}. The future Cherenkov Telescope Array (CTA) will be about 10 times more sensitive than H.E.S.S.\ and is expected to see TeV emission from a few more GCs. GCs have also been detected in radio (e.g., \cite{Clapson2011}) and diffuse X-rays (e.g., \cite{Eger2010,Eger2012,Wu2014}).  

Our motivation is to study the detectability of 16 Galactic GCs\footnote{We decided to revisit the 15 sources selected by Abramowski et al.~\cite{Abramowski2013} as well as Terzan~5, because new data and updated analysis methods have come available since that study was published.} for H.E.S.S.\ and CTA, and to rank them according to their predicted TeV flux. In Section~\ref{model}, we briefly discuss the model of Kopp et al.~\cite{Kopp2013},  after which follows excerpts of a parameter study to investigate the model's behaviour and to study the degeneracy between free parameters (Section~\ref{param}). In Section~\ref{analysis} we present the results from reanalysing H.E.S.S.\ data on GCs. Section~\ref{cons} includes a discussion on the parameters of Terzan~5 that we have constrained using multi-wavelength data; we then list the five most promising GCs for CTA based on their predicted VHE flux in Section~\ref{SEDs}. Our conclusions follow in Section~\ref{conclude}. For more details, see Ndiyavala et al., \emph{submitted}.

\section{The leptonic transport and emission model for GCs}
\label{model}
We used a multi-zone, steady-state, spherically symmetric model~\cite{Kopp2013}, that assumes pulsars are sources of relativistic leptons in GCs to calculate the particle transport (including diffusion and radiation losses) and to predict the spectral energy distribution (SED) expected from GCs for a very broad energy range by considering synchrotron radiation (SR), as well as inverse Compton (IC) emission. The Fokker-Planck type equation in Parker~\cite{Parker1965} prescribes the transport of charged energetic particles, i.e., electrons and positrons. Neglecting spatial convection, it is given by:
\begin{equation}
 \frac{\partial n_{\rm e}}{\partial t} = \kappa \nabla^{2} n_{\rm e} - 
\frac{\partial}{\partial E_{\rm e}}(\dot{E}_{\rm e}n_{\rm e}) + Q,
\end{equation}
where $n_{\rm e}$ is the electron density (per energy and volume) as a function of central radius ${r}_{\rm s}$ and particle energy $E_{\rm e}$; $\kappa$ is the diffusion coefficient, $\dot{E}_{\rm e}$ denotes the radiation losses, and $Q$ is the source term. In order to calculate the IC losses $\dot E_{\rm IC}$, we consider blackbody soft-photon densities \cite{Zhang2008} due to the cosmic microwave background (CMB) and photons from stars with a temperature of $T_{1} = 4~500~\rm K$. For the stellar photons, we used the line-of-sight integral for the photon number density  \cite{Prinsloo2013,Kopp2013},
\begin{equation}\nonumber
n_{\varepsilon, j}(r_{\rm s}, \varepsilon, T_{1}) = \frac{8\pi}{h^{3}c^{3}}\frac{\varepsilon^{2}}{e^{\frac{\varepsilon}{k_{\rm 
B}T_{1}}}-1}\left(\frac{1}{2}\frac{N_{\rm tot}R_{\star}^{2}}{R_{\rm c}^{2}\widetilde{R}}\right) 
\int_{r^{\prime}=0}^{r^{\prime}=R_{t}}\hat\rho(r^{\prime})\\
\frac{r^{\prime}}{r_{\rm s}}~{\rm ln}~\left(\frac{\mid r^{\prime} + r_{\rm s}\mid}{\mid r^{\prime} - 
r_{\rm s}\mid}\right)\,dr^{\prime},
\label{eq:16} 
\end{equation}
where~$N_{\rm tot}$~represents the total number of cluster stars, which can be written as $N_{\rm tot}={M_{\rm tot}}/{\-m}$, with~$M_{\rm tot}$ the total mass of the cluster and~$\-m$ the average stellar mass. Here, $R_{\star}$~is the 
average stellar radius, $R_{\rm c}$~indicates the core radius\footnote{The core radius is the distance from the centre of the cluster at which the apparent surface brightness of the cluster reduces by half.} of the cluster, and~$\widetilde{R} = 2R_{\rm h}-2R_{\rm c}/3 - {R_{\rm h}^{2}}/{R_{\rm r}}$, with $R_{\rm h}$ the half-mass radius\footnote{The half-mass radius is the radius from the core including half of the total mass of the cluster.} and $R_{\rm t}$~the tidal radius\footnote{The tidal radius is the distance from the cluster core beyond which the gravitational influence of the Galaxy is larger than that of the GC.} of the cluster. 

In the case of SR, we assumed a constant $B$-field to calculate the SR radiation losses. We considered Bohm diffusion
\begin{eqnarray}
\kappa(E_{\rm e}) & = & \frac{c}{3e}\kappa_{B}\frac{E_{\rm e}}{B}, 
\end{eqnarray}
with $c$ being the speed of light and $e$ the elementary charge. We also investigated diffusion coefficients of the form $\kappa(E_{\rm e}) = \kappa_B(E_{\rm e}/E_0)^{\alpha}$ where $E_{0}=1~\rm TeV~\rm and~\alpha=0.6$ (e.g., \cite{Moskalenko1998}). Lastly we used a power-law particle injection spectrum:
\begin{eqnarray}
 Q(E_{\rm e}) & = & Q_{0}E_{\rm e}^{-\Gamma}
\end{eqnarray}
between energies $E_{\rm e, \rm min}=100$~GeV and~$E_{\rm e, \rm max}$ is assumed to be $\leq100$ TeV. The value for the source strength or normalisation $Q_{0}$ were obtained using
\begin{eqnarray}
 \int E_{\rm e}Q(E_{\rm e})\,dE_{\rm e} & = & \eta \langle \dot {E} \rangle N_{\rm MSP},
 \end{eqnarray}
with $\eta$ the particle conversion efficiency i.e., the fraction of pulsar spin-down power that is converted to particle power, $N_{\rm MSP}$ the number of MSPs in the GC, and $\langle\dot{E}\rangle$ the average MSP spin-down power.

\section{Parameter study} 
\label{param}
\begin{figure}[t]
\centering
\begin{minipage}{.75\textwidth}
  \centering
  \includegraphics[height=4.8 cm, width=8.5cm]{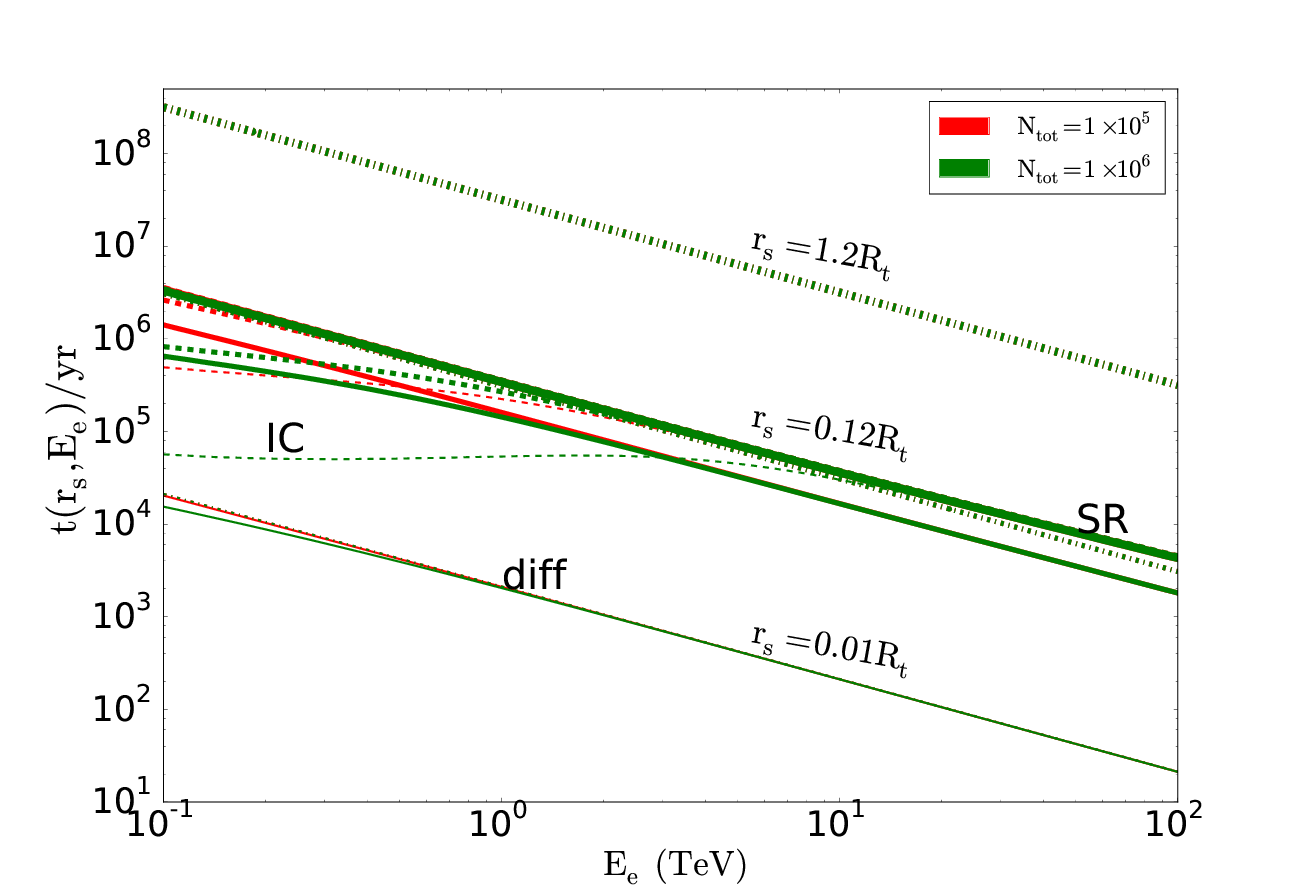}
  \caption{Timescales as a function of energy for diffusion (dotted lines), 
radiation losses (dashed lines), and the effective scale (solid lines). Thicker 
lines represent larger radii. The colours represent different values of $N_{\rm tot}$ 
as noted in the legend.}
  \label{fig:tls_stars_energy}
\end{minipage}%
\end{figure}
We have performed an extensive parameter study using our code. As an example, we present the effect of changing $N_{\rm tot}$ here. Figure~\ref{fig:tls_stars_energy} presents different timescales versus energy: Radiation
($\tau_{\rm rad}^{-1}=\tau^{-1}_{\rm IC}+\tau^{-1}_{\rm SR}$, dashed lines, where $\tau_{\rm IC}=E_{\rm e}/\dot {E}_{\rm e, IC}$ and $\tau_{\rm SR}=E_{\rm e}/\dot {E}_{\rm e, SR}$), escape ($\tau_{\rm diff}= r^{2}/(2\kappa)$; dotted lines), and effective timescale ($\tau_{\rm eff}^{-1}=\tau^{-1}_{\rm rad}+\tau^{-1}_{\rm diff}$); solid lines). The line thickness indicates different radii. The number density of stellar soft photons $n_{\varepsilon}$ scales 
linearly with $N_{\rm tot}$ (see Eq.~[\ref{eq:16}]). Therefore, so does the IC loss rate associated with the stellar component. This can be seen in Figure~\ref{fig:tls_stars_energy} at low energies.
For a smaller $N_{\rm tot}$, $n_\varepsilon$ is lower and hence the IC loss rate drops. It therefore takes a longer time for the particles to lose energy. The IC cross section also drops as one goes from the Thomson regime at low energies to the Klein-Nishina regime at high energies. Therefore, the SR loss rate determines the effective timescale at high energies. At the smallest radii, diffusion dominates over radiation losses. At intermediate radii one can see the change in regime: for $r_{\rm s}= 0.12R_{\rm t}$, with $R_{\rm t}$ the tidal radius, the SR timescale is only slightly lower than the diffusion timescale at the highest particle energies, and therefore determines the effective timescale in this case.
At larger radii, $n_\varepsilon$ rapidly declines (leading to smaller $\dot{E}_{\rm IC}$~and longer $\tau_{\rm IC}$) and SR losses dominate over diffusion (particle escape). 

\begin{figure}[!h]
\centering
\begin{minipage}{.75\textwidth}
  \centering
  \includegraphics[height=4.8cm, width=8.5cm]{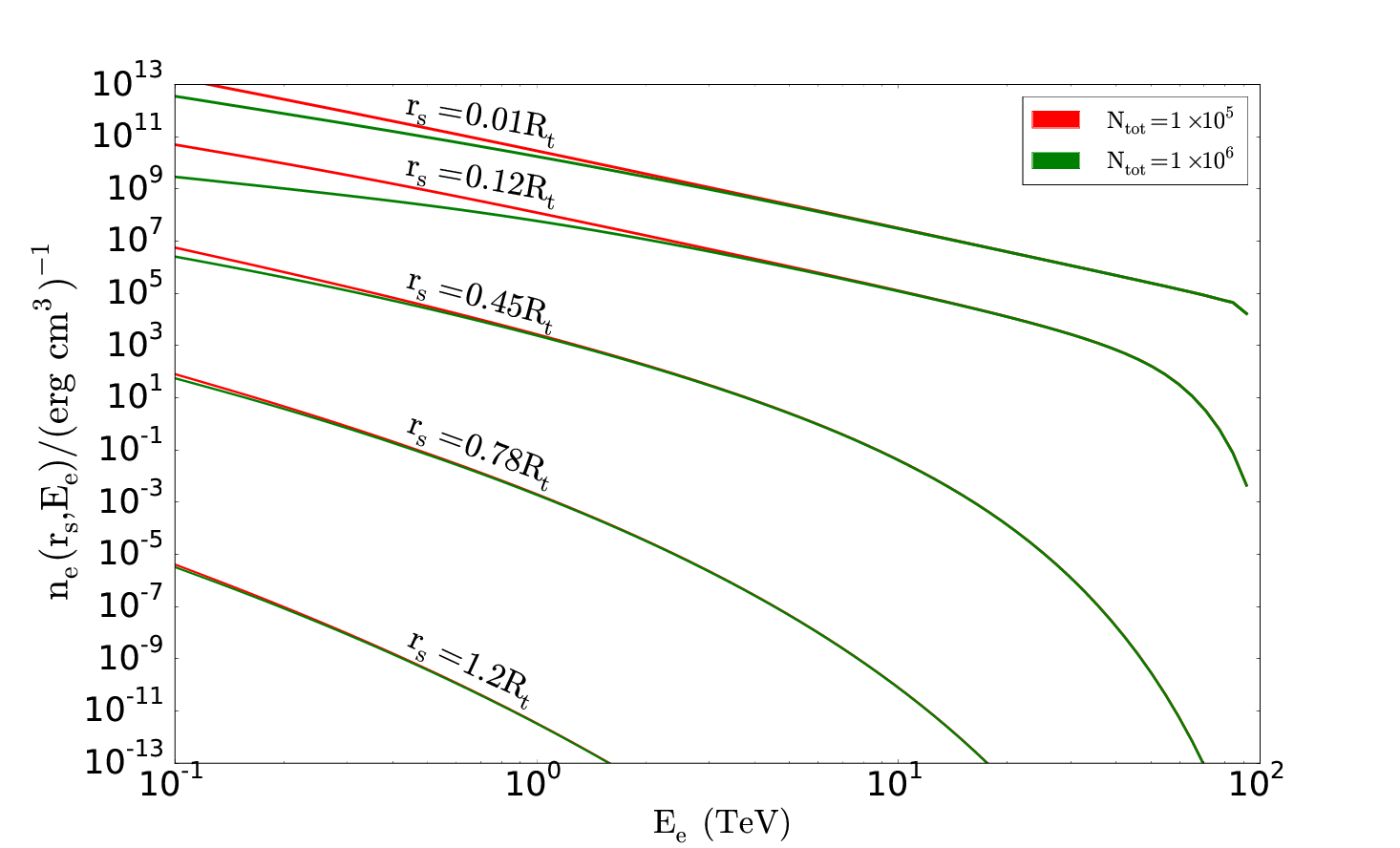}
  \caption{The steady-state particle spectrum as a function of particle energy $E_{\rm e}$ for different radii $r_{\rm s}$.}
  \label{fig:mel_stars_energy}
\end{minipage}
\end{figure}

In Figure~\ref{fig:mel_stars_energy}, at a fixed radius, the steady-state particle spectrum $n_{\rm e}$ is higher for a smaller value of $N_{\rm tot}$ (at low energies). This is because $\dot E_{\rm IC}$ is lower in this case. At large 
energies, this effect vanishes because SR cooling dominates and it is not a function of $N_{\rm tot}$. At larger radii the effect of changing $N_{\rm tot}$ on the value of $n_{\rm e}$ is smaller,  because $n_{\varepsilon}$ and therefore  $\dot E_{\rm IC}$ 
decreases rapidly with distance. One can see that there is a cutoff at higher energies due to SR. The cutoff energy becomes increasingly lower at larger radii since high-energy particles continue to lose energy due to SR. Furthermore, the overall level of $n_{\rm e}$ decreases with radius since it represents a particle density, and the volume scales as $r_{\rm s}^{3}$. 

\section{Reanalysis of H.E.S.S.\ data}
\label{analysis}
H.E.S.S.\ searched for VHE signals from 15 GCs in their archival data~\cite{Abramowski2013} since many GC positions were covered by the H.E.S.S.\ Galactic Plane Survey \cite{Gast2011} or lay in the same field of view (FoV) of other observed H.E.S.S\ sources. The GC catalogue of \cite{Harris2010} was used to select the 15 GCs which lay within  $1.0^\circ$ of the Galactic Plane~\cite{Abramowski2013}. The data runs furthermore should have passed the standard quality selection criteria. H.E.S.S.\ saw no significant excess emission above the estimated background for any of the 15 selected GCs.
H.E.S.S.\ has accumulated more data since the previous analysis, and thus we decided to reanalyse the H.E.S.S.\ data to investigate whether we could find deeper flux upper limits which would be more constraining to our GC emission model. We compared our new results with those of the prior study \cite{Abramowski2013}, and found that our results on the 15 GCs were fully consistent with the earlier ones (we could not significantly detect any of the 15 GCs). We therefore decided to use the earlier published results to constrain our model parameters in what follows. We also performed a stacking analysis to search for a population of faint emitters. The total GC stack had an acceptance-corrected livetime of 644 hours of good quality data. Our new stacking upper limit was consistent with the published one~\cite{Abramowski2013}. We lastly studied Terzan~5, which is the only GC that has been plausibly detected at a significance of 5.3$\sigma$ in the VHE band \cite{Abramowski2011}. During our reanalysis of H.E.S.S.\ data, we confirmed the detection of Terzan~5 at a similar significance level.

\section{Constraining model parameters via X-ray and gamma-ray data}
\label{cons}
\begin{figure}[t]
\centering
\begin{minipage}{.7\textwidth}
  \centering
 \includegraphics[height=5.0cm, width=8.5cm]{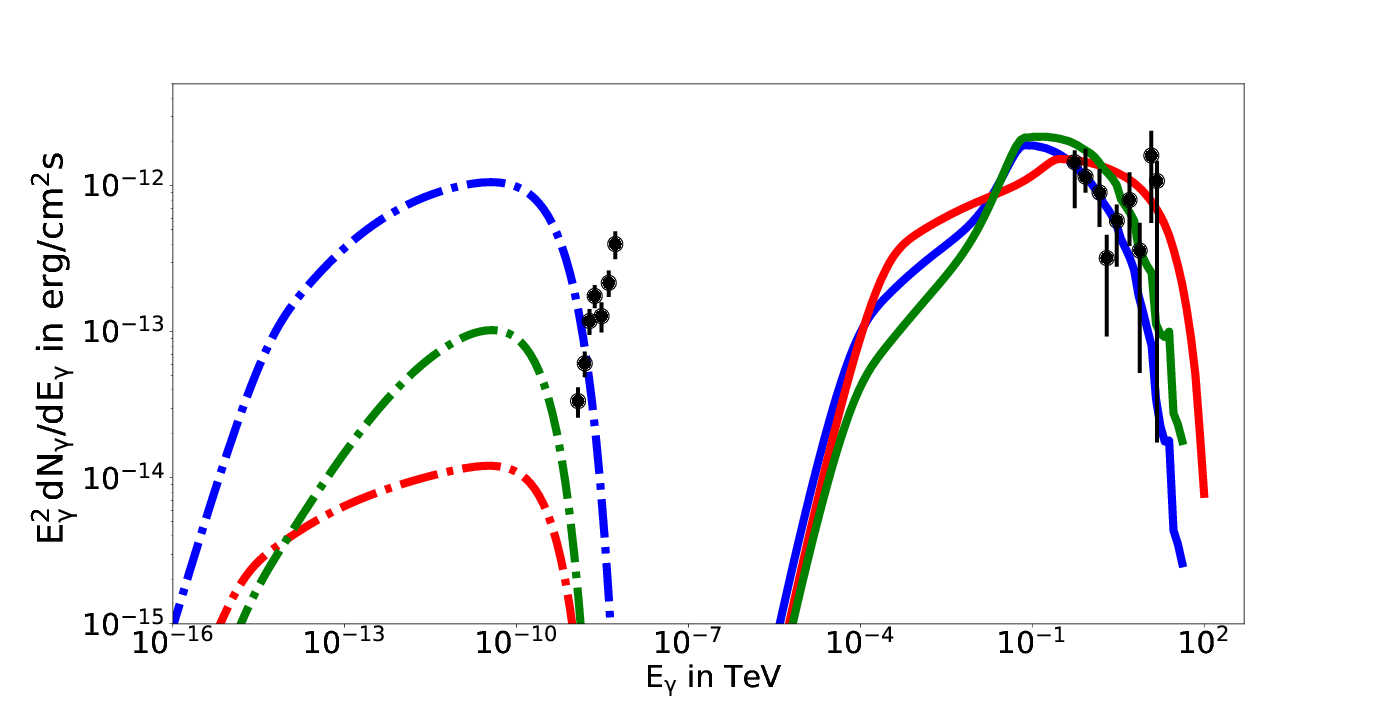}
\caption{The SED for Terzan~5 indicating the predicted SR (integrated between $55^\prime<r_{\rm s}<174^\prime$ to match \textit{Chandra} FOV, the dash-dotted lines) for the inner part of the source and IC (integrated over all $r_{\rm s}$) components using combinations of parameters which do not violate the \textit{Chandra} \cite{Eger2010} and H.E.S.S.\ \cite{Abramowski2011} data.}
 \label{terzan5}  
\end{minipage}%
\end{figure}

We used diffuse \emph{Chandra} X-ray and H.E.S.S.\ VHE gamma-ray observations to constrain cluster parameters for three sources ( i.e., Terzan~5, 47~Tucanae, and NGC~6388) so as not to violate the data. As an example, we present the results for Terzan~5, using the structural parameters given in Table~\ref{GCslist}. Our model cannot reproduce the flat slope of the X-ray data. Hence, we postulate a new radiation component (see Venter et al., \emph{in preparation}, who attribute this to cumulative pulsed SR from the individual MSP magnetospheres) to explain these data. We therefore treat the X-ray data as upper limits and our predicted SR component must lie below these. Figure~\ref{terzan5} shows the predicted differential SED components of Terzan~5 (with gamma-ray \cite{Abramowski2011} and X-ray \cite{Eger2010} data overplotted) using three combinations of parameters: the blue lines represent the case for Bohm diffusion, $B = 5\,\mu\rm{G},\,\Gamma = 1.8,\,Q_{0} = 1.16\times10^{34}{\rm erg^{-1}s^{-1}},\,E_{\rm e,max} = 100\,TeV$; the red line represents the case for Bohm diffusion, $B = 1\,\mu\rm{G},\,\Gamma = 1.8,\,Q_{0} = 6.33\times10^{33}{\rm erg^{-1}s^{-1}}\,E_{\rm e,max} = 20\,TeV$; and the green line represents the case for $\kappa_{0} = 0.7\times10^{-4}\, {\rm kpc^{2}/Myr},\,B = 2\,\mu\rm{G},\,\Gamma = 2,\,Q_{0} = 9.84\times10^{33}{\rm erg^{-1}s^{-1}},{\rm and}\,E_{\rm e,max} = 50\,TeV$. We see that there are different parameters combinations that satisfy the observational constraints, indicating degeneracy between model parameters and the need for more low-energy data.

\section{Ranking the GCs according to predicted VHE flux}
\label{SEDs}
\begin{figure}[!h]
\centering
\includegraphics[height=5.5 cm, width=9.0cm]{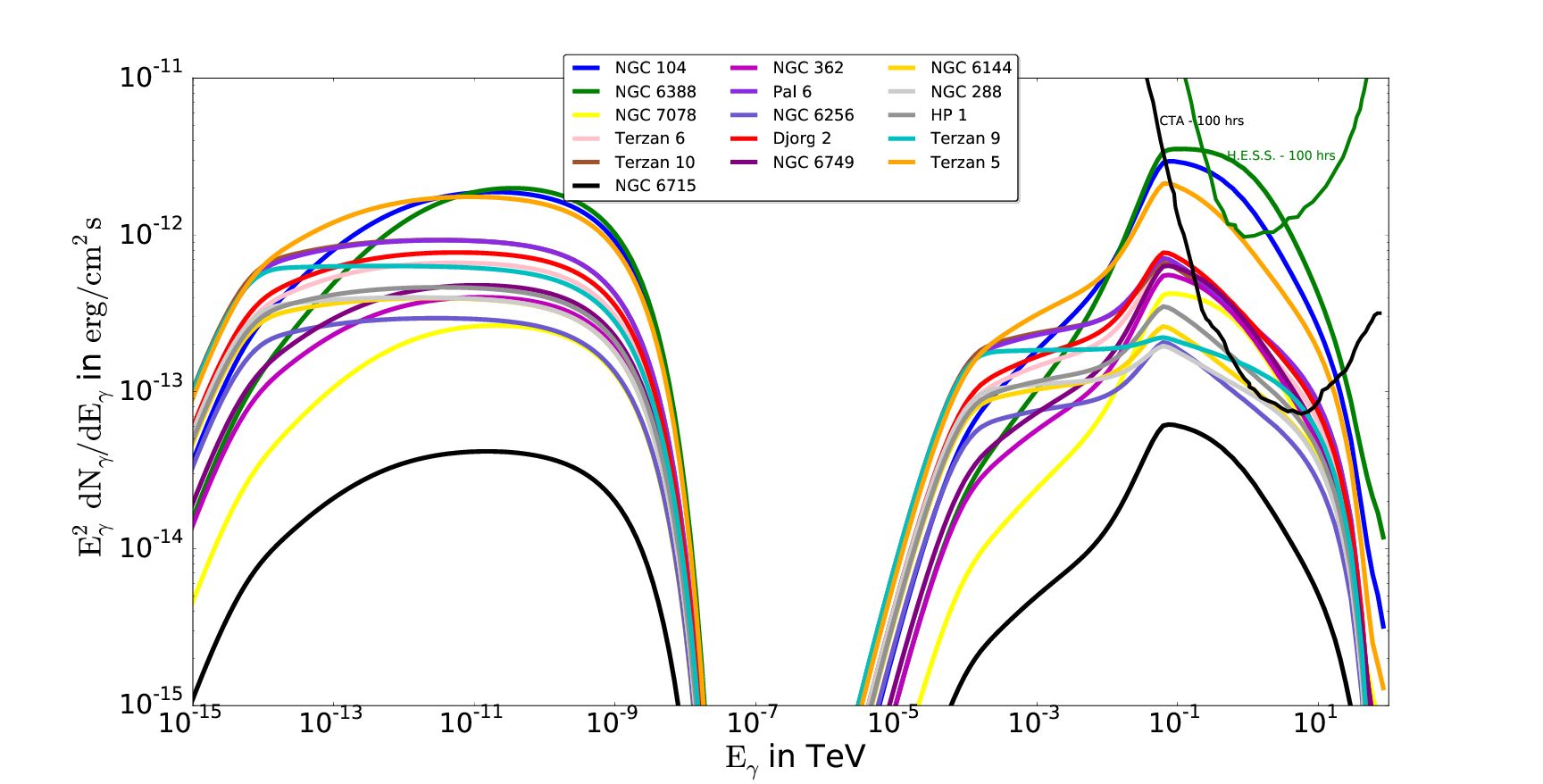}
\caption{Predicted SED ${E^{2}_{\gamma}} \, {dN_{\gamma}/dE_{\gamma}}$ in ${\rm erg\,cm^{-2}s^{-1}}$ for 15 non-detected GCs and for Terzan~5. The two components represent the SR and IC spectra. The H.E.S.S.\ and CTA sensitivities (for 100 hours) are also shown.}
\label{GCs}
\end{figure}

We applied the model described in Section~\ref{model} to 15 non-detected GCs at TeV energies and to Terzan~5 using fixed parameters (see Table~\ref{GCslist}). 
We have used typical values for $N_{\rm MSP}$, e.g.,~\cite{Abdo2010,Venter2015}, and $N_{\rm tot}$ values from Lang~\cite{Lang1992} and obtained distances $d$ and structural parameters from Harris~\cite{Harris1996,Harris2010}. We assumed Bohm diffusion, $\Gamma = 2.0$, and $B = 5 \mu G$ to produce SR and IC spectra for each individual cluster. From Figure~\ref{GCs} we can see that H.E.S.S.\ may possibly detect three GCs, i.e., Terzan~5 (orange),
47~Tucanae (blue), and NGC 6388 (green) if the telescope observes these sources for 100 hours. 47~Tucanae and NGC~6388 are currently not detected by H.E.S.S.; they were only observed for about 20 hours each. We note, however, that this flux prediction and therefore ranking is very sensitive to the choice of parameters. The CTA will be 10 times more sensitive than H.E.S.S.\ and should therefore detect many more GCs (we find that more than half of the known Galactic population may be detectable, depending on observation time and model parameters). The five most promising GCs for CTA observations are NGC~6388, 47~Tucanae, Terzan~5, Djorg~2, and Terzan~10. 

\begin{table}[t]
\centering
\small
  \begin{tabular}{|l|r|r|r|r|r|r|r|}
\hline
GC name  & \vtop{\hbox{\strut $d$} \hbox{\strut (kpc)}}  & 
\vtop{\hbox{\strut $N_{\rm tot}$} \hbox{\strut($10^{5}$)}} & 
\vtop{\hbox{\strut$N_{\rm MSP}$}} &
\vtop{\hbox{\strut $Q_{0}$} \hbox{\strut($10^{33}\rm {/erg/s})$}} &
\vtop{\hbox{\strut$R_{\rm c}$} \hbox{\strut $(^\prime)$}} &
\vtop{\hbox{\strut$R_{\rm h}$} \hbox{\strut $(^\prime)$}}&
\vtop{\hbox{\strut$R_{\rm t}$} \hbox{\strut $(^\prime)$}} \\

 \hline
 47~Tucanae    & 4.5  & 4.57 & 33  & 9.55  & 0.36 & 3.17 & 42.86   \\
 NGC 6388   & 9.9  & 5.81 & 180 & 52.1  & 0.12 & 0.52 & 6.21  \\ 
 NGC7078    & 10.4 & 4.13 & 25  & 7.24  & 0.14 & 1.00 & 21.5   \\
 Terzan 6   & 6.8  & 0.29 & 25  & 7.24  & 0.05 & 0.44 & 17.39   \\
 Terzan 10  & 5.8  & 0.38 & 25  & 7.24  & 0.9  & 1.55 & 5.06  \\
 NGC 6715   & 26.5 & 4.79 & 25  & 7.24  & 0.09 & 0.82 & 7.47   \\
 NGC 362    & 8.6  & 1.58 & 25  & 7.24  & 0.18 & 0.82 & 16.11   \\
 Pal 6      & 5.8  & 0.31 & 25  & 7.24  & 0.66 & 1.2  & 8.36  \\
 NGC 6256   & 10.3 & 0.21 & 25  & 7.24  & 0.02 & 0.86 & 7.59   \\
 Djorg 2    & 6.3  & 0.51 & 25  & 7.24  & 0.33 & 1.0  & 10.53   \\
 NGC 6749   & 7.9  & 0.24 & 25  & 7.24  & 0.62 & 1.1  & 5.21  \\
 NGC 6144   & 8.9  & 0.48 & 25  & 7.24  & 0.94 & 1.63 & 33.25   \\
 NGC 288    & 8.9  & 0.32 & 25  & 7.24  & 1.35 & 2.23 & 12.94 \\
 HP 1       & 8.2  & 0.48 & 25  & 7.24  & 0.03 & 3.1  & 8.22  \\
 Terzan 9   & 7.1  & 0.02 & 25  & 7.24  & 0.03 & 0.78 & 8.22   \\
 Terzan~5   & 5.9  & 8.0  & 34  & 6.33  & 0.10 & 0.72 & 13.27  \\
\hline
\end{tabular}
\caption{In this table we list structural parameters of the 16 GC. The first 15 parameters is taken from Table 1 in Venter et al.~\cite{Venter2015} and the parameters of Terzan~5 is taken from Harris~\cite{Harris2010}. The columns are cluster identification; distance in kpc; estimated number of stars~\cite{Lang1992}; number of $\rm MSPs$; source strengths $Q_{0}$; core radius; tidal radius; and half-mass radius.}
\label{GCslist}
\end{table}

\section{Conclusion}
\label{conclude}
We have briefly described an emission model that we applied to 15 GCs that have been observed, but not detected, in VHE gamma rays, as well as to Terzan~5. While the parameters of the individual GCs are uncertain (and sometimes degenerate), we noted that most of the flux predictions for the GCs are below the H.E.S.S.\ sensitivity limit, but that CTA may detect many more GCs (possibly tens of sources) because it will be 10 times more sensitive than H.E.S.S. Future multi-wavelength studies should allow us to constrain some parameters as well as discriminate between competing radiation models. 

\ack
This work is based on the research supported wholly in part by the National Research Foundation of South Africa (NRF; Grant Numbers 93278, and 99072). The Grantholder acknowledges that opinions, findings and conclusions or recommendations expressed in any publication generated by the NRF supported research is that of the author(s), and that the NRF accepts no liability whatsoever in this regard.
\section*{References}
\bibliography{iopart-num}
\end{document}